\documentclass[useAMS]{mn2e}
\usepackage{mnras_cite}
\usepackage{epsfig}

\def\G{{\rm G}}

\def\H2{{\rm H}$_2$}

\def\d{{\rm d}}

\begin{document}

\title[Tidal Mass Loss from Collisionless Systems]{Tidal Mass Loss from Collisionless Systems}
\author[Marios Kampakoglou, Andrew~J.~Benson]
{Marios Kampakoglou
and Andrew~J.~Benson\thanks{Current address: California Institute of Technology, MC 130-33, 1200 E. California Blvd., Pasadena CA 91125, U.S.A. (e-mail: abenson@its.caltech.edu)} \\
Department of Physics, University of Oxford, Keble Road, Oxford OX1 3RH, United Kingdom (e-mail: mariosk,abenson@astro.ox.ac.uk)\\
}

\maketitle

\begin{abstract}
We examine the problem tidally-induced mass loss from collisionless systems such as dark matter haloes. We develop a model for tidal mass loss, based upon the phase space distribution of particles, which accounts for how both tidal and Coriolis torques perturb the angular momentum of each particle in the system. This allows us to study how both the density profile and velocity anisotropy affect the degree of mass loss---we present basic results from such a study. Our model predicts that mass loss is a continuous process even in a static tidal field, a consequence of the fact that mass loss weakens the potential of the system making it easier for further mass loss to occur. We compare the predictions of our model with N-body simulations of idealized systems in order to check its validity. We find reasonable agreement with the N-body simulations except for in the case of very strong tidal fields, where our results suggest that a higher-order perturbation analysis may be required. The continuous tidally-induced mass loss predicted by our model can lead to substantial reduction in satellite mass in cases where the traditional treatment predicts no mass loss. As such, our results may have important consequences for the orbits and survival of low mass satellites in dark matter haloes.
\end{abstract}

\section{Introduction}

It is well known that our Galaxy is surrounded by a family of satellite systems. These include the several known dwarf spheroidal
galaxies as well as the Large and Small Magellanic Clouds and numerous globular clusters \cite{mateo98}. The gravitational tidal
influence of our Galaxy is currently causing the Sagittarius galaxy and globular cluster Palomar 51 to be torn apart
\cite{ibata94,odenkirhen01}. Tidally induced mass loss may also be important for explaining the amount and distribution of
intracluster light \cite{calcaneo00,zwicky51}. An improved understanding of the way in which tidal mass loss occurs is therefore
of great importance for understanding the evolution of galaxies and clusters.

Furthermore it is known that, although N-body simulations of cold dark matter (CDM) successfully reproduce the overall structure of
the Universe, they show potential discrepancies with observations at galactic scales. The discrepancy between the number of low mass sub-haloes predicted by CDM and the observed number of satellite galaxies is on the order of a factor 5--10 more \cite{moore99,tasitsiomi02}. This is the so-called satellite
catastrophe for hierarchical galaxy models.

Tidal interactions between galaxies may help to resolve this apparent discrepancy \cite{bullock00,benson02}. Sub-haloes can fail
to be ``lit-up'' by a galaxy because their star formation has been inhibited by early and significant mass loss. There is also the
possibility, in extreme cases, for the satellite to be totally destroyed due to the gravitational influence of a larger dark matter
halo or massive galaxy.

Finally, observations show that merging is a key driver in the evolution of galaxy morphologies. N-body simulations of the
assembly of galaxies in a hierarchical Universe show that, as dark matter halos coalesce, the embedded galaxies merge on a
time-scale that is consistent with dynamical friction estimates based on their total mass \cite{navarro95}. Accurate treatment
of tidal interactions is important if we wish to explore this phenomena since dynamical frictional timescales depend upon the
remaining bound mass of each satellite.

Recently, analytic models of satellite orbits and merging have been developed and employed to study the properties of the sub-halo
populations of CDM haloes \cite{taylor01,benson02,benson05,penarrubia04,just04}, and
to make quantitative predictions for the distributions of such sub-haloes. These models typically employ what we will refer to as the ``classic'' treatment
of tidal mass loss---namely computing a radius in the satellite at which internal and tidal forces balance and removing all mass beyond that
radius---together with simple estimates of the time over which this mass is lost. This ``classic'' method of determining tidal
mass loss is clearly oversimplified---as particles orbit within the satellite halo they may be stripped away by tidal forces as
they approach apocentre, even though they spend most of their time at smaller radii. In this work, therefore, we re-examine the
process of tidal mass loss and present a more realistic calculation based upon the orbits of individual particles within a
satellite.

The layout of this paper is as follows. Section~\ref{sec:alyt} briefly reviews the ``classic'' calculation of tidal radius, before
describing our improved model and presenting some basic results from it. Section~\ref{sec:nbody} compares the results of our model with N-body simulations. Finally, in \S\ref{sec:discuss} we
discuss our results and present our conclusions.

\section{Analytic Calculation of Post-Tidal Density Profile}
\label{sec:alyt}

We begin by reviewing the classic calculation of tidal mass loss in \S\ref{sec:classcalc}, before presenting our improved calculation in \S\ref{sec:phspmod}.

\subsection{Classical Calculation}
\label{sec:classcalc}

In order to determine the tidal radius, we consider a small mass orbiting at radius $r$ from the centre of a satellite and let
the mass of satellite enclosed inside this orbit be $M_{\rm s}(<r)$. The satellite itself orbits within a host potential at a radius
$r^\prime$. There are two forces on this mass: the gravitational pull towards the centre of the satellite and the
gravitational tidal force from the host potential.

The gravitational acceleration on the mass towards the satellite centre is given by
\begin{equation}
a_{\rm s} = -{\G M_{\rm s} \over r_{\rm s}^2} {y_{\rm s}(x) \over x^2},
\end{equation}
where $x \equiv r/r_{\rm s}$, $y_{\rm s}(x) \equiv M_{\rm s}(<r)/M_{\rm s}$, and $M_{\rm s}$ and $r_{\rm s}$ are a characteristic
mass and radius respectively in the satellite. The gravitational tidal acceleration on the mass due to the host potential can
be expressed as:
\begin{equation}
a_{\rm t} = x  {\G M_{\rm s} \over r_{\rm s}^2} {\d \over \d x^\prime} {y^\prime(x^\prime) \over x^{\prime 2}},
\end{equation}
where $x^\prime \equiv r^\prime/r_{\rm s}$, $y^\prime(x^\prime) \equiv M^\prime(<r^\prime)/M_{\rm s}$, and $M^\prime(<r^\prime)$
is the mass of the host system within radius $r^\prime$. Note that we have approximated the tidal field as being linear across the
satellite---the usual approximation made in considering tidal fields. The tidal radius is reached when the gravitational acceleration on the mass towards the satellite centre equals the tidal
acceleration on the mass from the host potential, i.e. where $a_{\rm s} = a_{\rm t}$ or
\begin{equation}
\frac{y_{\rm s}(x)}{x^{2}} = - x {\d \over \d x^\prime} {y^\prime(x^\prime) \over x^{\prime 2}}.
\end{equation}

\begin{figure}	
\epsfig{file=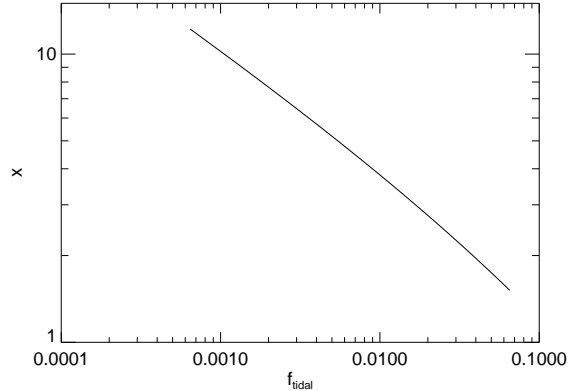,width=8cm}
\caption{The tidal radius in the classical model plotted as function of the tidal field strength, $f_{\rm tidal}$, for the satellite galaxy we assume an NFW halo profile with concentration parameter $c_{\rm NFW}=11$.}
\label{fig:classrad}
\end{figure}

The above equation can be re-expressed in terms of the tidal field $f_{\rm tidal}$, defined to be the gradient of the tidal
acceleration due to the host potential
\begin{equation}
f_{\rm tidal}= - {\d \over \d x^\prime} {y^\prime(x^\prime) \over x^{\prime 2}} = c_{\rm tidal} {y^\prime(x^\prime) \over x^{\prime 3}},
\end{equation}
where
\begin{equation}
c_{\rm tidal}= -{x^{\prime 3}\over y^\prime(x^\prime) }{\d \over \d x^\prime} {y^\prime(x^\prime) \over x^{\prime 2}},
\end{equation}
giving
\begin{equation}
\frac{y_{\rm s}(x)}{x^3}=f_{\rm tidal}.
\end{equation}
The default values we choose are typical of a massive satellite in a Milky Way-like system ($M_{\rm s}=10^{10} M_{\odot}, M^\prime=10^{12} M_{\odot}$ and a distance between the centre of mass of the satellite and that of the host galaxy of $R=100$kpc, resulting in $f_{\rm tidal}=5.8\times 10^{-3}$---we will explore variations of a factor of 10 about this default $f_{\rm tidal}$).
The tidal radius can be found by solving the above for $x$. Figure~\ref{fig:classrad} shows the tidal radius as a function of
$f_{\rm tidal}$ for a satellite described by an NFW density profile \cite{navarro95} with a concentration parameter (the ratio of the virial radius to the characteristic scale radius) of $c_{\rm
NFW}=11$. (For the characteristic radius and mass we have chosen $r_{\rm s}$ to be the scale radius of the NFW profile and have
set $M_{\rm s}$ to equal $M_{200}$---the virial mass of the halo.)

\subsection{Phase Space Calculation}
\label{sec:phspmod}

As mentioned in the Introduction, the classic model of tidal mass loss is clearly highly simplified. It assumes a sharp cut-off in
the satellite density profile at the tidal radius, and is insensitive to the phase space distribution of the satellite
(i.e. satellites in which orbits are purely tangential would have the same tidal radius as those with purely radial orbits,
providing that the overall density profile was the same). Furthermore the classic model does not take into account the orbit of stars, we know that prograde orbits are more easily stripped than radial orbits; while radial orbits are more easily stripped than retrograde orbits \cite{read06}.

\begin{figure}
\epsfig{file=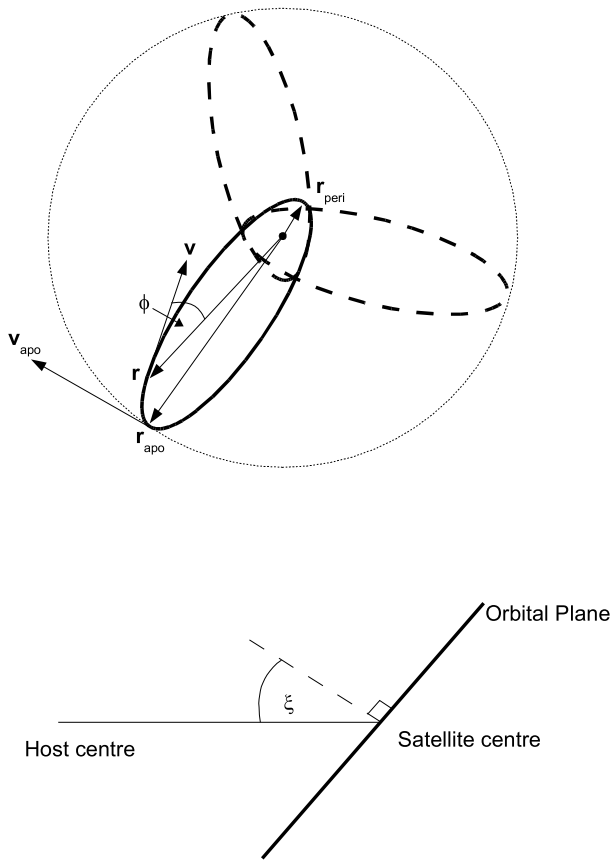,width=8cm}
\caption{The geometry of the satellite and particle orbit used in our calculation of mass loss. The upper diagram coincides with
the orbital plane. The particle is identified at position $\mathbf{r}$ at which it has velocity $\mathbf{v}$. The angle between
$\mathbf{r}$ and $\mathbf{v}$ is $\phi$. The orbit of the particle has pericentre and apocentre $\mathbf{r}_{\rm peri}$ and
$\mathbf{r}_{\rm apo}$ respectively. At apocentre, the particle velocity is $\mathbf{v}_{\rm apo}$. The orbit is assumed to
precess within the satellite potential (as schematically indicated by the dashed ellipses) and eventually sweeps out a region
indicated by the dotted circle. The diameter of this circle is $\Delta r$. The lower diagram coincides with the plane containing
the line which connects the satellite centre to the centre of the host potential, and the normal to the orbital plane. The angle
between these two lines is $\xi$.}
\label{fig:diag}
\end{figure}

In order to understand the effects of a tidal field on the particles (i.e. particles of dark matter or stars\footnote{Our
calculation applies to any collisionless particle and can also be trivially extended to composite systems, such as a stellar
disk within a dark matter halo.}) within the satellite with mass $M_{\rm sat}$, we assume a satellite which experiences a constant
tidal force (e.g. on a circular orbit within a spherical host potential with negligible dynamical friction). Although the assumption of a constant tidal force is unrealistic (few satellites will be on orbits that are close to circular and, for our default satellite, the dynamical friction timescale is of order 10Gyr, resulting in significant change in orbital radius on a timescale comparable to the mass loss timescale from the satellite), we choose to explore this idealized case here and will explore more realistic examples in future work. Particle orbits within the satellite however, have
arbitrary ellipticity. Figure~\ref{fig:diag} shows the geometry of the particle orbit considered below.

The energy per unit mass of a particle is
\begin{equation}
E(x) =\frac{1}{2}v^2 +{\Phi}(x),
\end{equation}
where ${\Phi}(x)$ is the gravitational potential of the satellite and $v$ is the velocity of the particle. We assume an NFW
potential for the satellite, resulting in a potential:
\begin{equation}
{\Phi}(x)=-4{\pi}\G\bar{\rho}r_{s}^2\frac{\ln(1+x)}{x},
\end{equation}
where $\bar{\rho}$ is the mean density of the halo.

We consider a rotating frame of reference which rotates at the same rate as the orbitting satellite. As the particle moves around its orbit it experiences a torque (relative to the centre of mass of the satellite) due to tidal (from the host potential) and
Coriolis forces. The angular momentum of the particle therefore varies around the orbit. The particle must always remain bound to
the combined satellite plus host potential (since those potentials are not time-varying the energy of the particle is conserved
along its orbit). However, we can ask the question: If the host potential were instantaneously removed, would the particle be
bound to the satellite or not? To answer this question we simply need to know the angular momentum of the particle relative to the
satellite centre, as this immediately gives the kinetic energy of the particle relative to the satellite centre. We expect this
criterion to be a reasonable discriminator between particles which are essentially confined to the satellite and those which can
move more freely throughout the host potential (i.e. those in tidal tails). Experiments with N-body simulations (see
\S\ref{sec:nbody}) confirm this expectation.

The calculation of the perturbation to the particle's angular momentum is given in Appendix~\ref{app:AM}, where we compute the maximum value of the particle's energy (in a frame instantaneously moving with the satellite), $E_{\rm max}$. The total mass loss
suffered from the satellite galaxy is found by integrating over phase space
\begin{eqnarray}
M_{\rm loss} & = & {{\int}_{0}^{x_{\rm max}}}{{\int}_{0}^\infty}{{\int}_{0}^{{\pi}}}{{\int}_{0}^{\frac{\pi}{2}}}4{\pi}r_{s}^3x^2F({\bf{x,v}})2{\pi}v^2{\sin}{\phi}{\sin}{\xi} \nonumber \\
 & & \times H(E_{\rm max}) \d{\xi} \d{\phi} \d v \d x.
\end{eqnarray}
For this work we assume that the phase space density, $F(\bf{x,v})$, is a separable function, so it can be written as:
\begin{eqnarray}
M_{\rm loss} & = & {{\int}_{0}^{x_{\rm max}}}{{\int}_{0}^\infty}{{\int}_{0}^{{\pi}}}{{\int}_{0}^{\frac{\pi}{2}}}4{\pi}r_{s}^3{\rho(x)}x^2f(v,\phi)2{\pi}v^2{\sin}{\phi}{\sin}{\xi} \nonumber \\
 & & \times H(E_{\rm max}) \d{\xi} \d{\phi} \d v \d x 
\label{eq:mloss}
\end{eqnarray}
where $\rho(x)$ is the density profile of the satellite and $H(s)=1$ if $s>0$ and $H(s)=0$ if $s<0$. Here, $f(v,\phi)$ is the velocity distribution function of the particles in the
satellite for which we assume an (anisotropic in general) Gaussian distribution and which is normalised such that $\int_0^\infty
\int_0^\pi f(v,\phi) 2 \pi v^2 \sin\phi \d\phi \d v=1$. The radial density profile can be found in a similar way as shown in eqn.~\ref{eq:mloss}, by simply not performing
the integral over radius.

\begin{figure}	
\epsfig{file=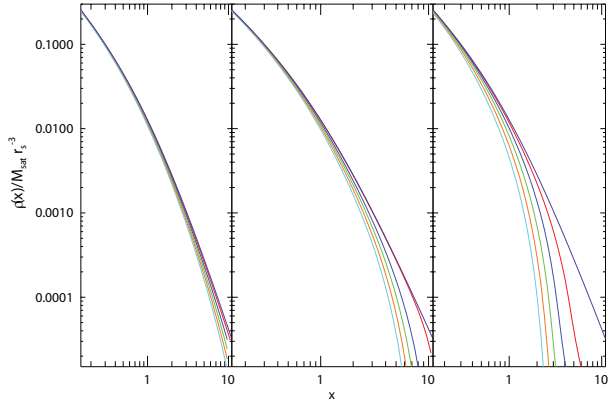,width=8cm}
\caption{The evolution of the mass density distribution of the satellite according to the model described in \S\ref{sec:phspmod}. The results are shown after 1, 2, 3, 4 and 5 mass loss iterations, and for different values of the tidal field: $f_{\rm tidal}=5.8\times10^{-4}$ (left panel), $5.8\times10^{-3}$ (middle panel) and $5.8\times10^{-2}$ (right panel). The initial---prior to stripping---satellite profile ({\it purple} line) is given for reference. Results are shown for an NFW satellite with concentration parameter $c=11$ and velocity anisotropy parameter ${\beta}=0$.}
\end{figure}

\begin{figure}	
\epsfig{file=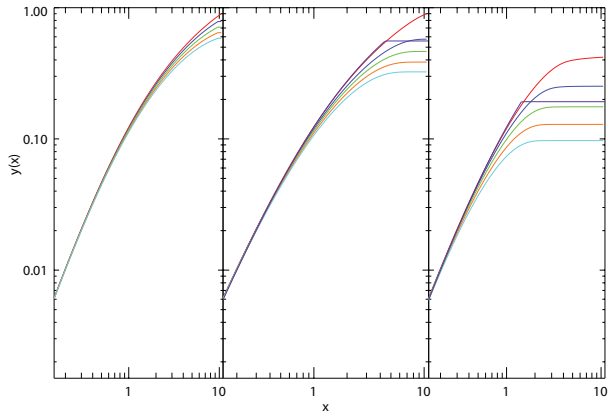,width=8cm}
\caption{The evolution of the satellite's bound mass distribution according to the model described in \S\ref{sec:phspmod}, where $y(x)=M_{\rm bound}(<x)/M_{\rm s}$, with $M_{\rm bound}(<x)$ being the remaining bound mass of the satellite within radius $x$. The results are shown after 1, 2, 3, 4 and 5 mass loss iterations, and for different values of the tidal field: $f_{\rm tidal}=5.8\times10^{-4}$ (left panel), $5.8\times10^{-3}$ (middle panel) and $5.8\times10^{-2}$ (right panel). Results are shown for an NFW satellite with concentration parameter $c=11$ and velocity anisotropy parameter ${\beta}=0$. In the middle and right-hand panels the purple line corresponds to the predictions from the classical approach. This method predicts zero mass loss for the plot in the left panel as the tidal radius is greater than the virial radius of the satellite.}
\label{fig:varftidal}
\end{figure}

\begin{figure}	
\epsfig{file=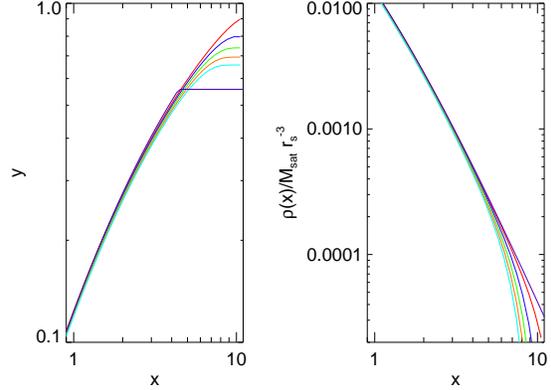,width=8cm}
\caption{The evolution of the satellite's bound mass distribution (left panel) and mass density distribution (right panel) according to the model described in \S\ref{sec:phspmod} for a velocity anisotropy parameter ${\beta}=0.6$. The purple line on the left panel corresponds to the predictions from the classical approach, on the right panel the initial---prior to stripping---satellite profile ({\it purple} line) is given for reference. The results are shown after 1, 2, 3, 4 and 5 mass loss iterations. The tidal field strength is $f_{\rm tidal}=5.8\times 10^{-3}$. Results are shown for an NFW satellite with concentration parameter $c=11$.}
\end{figure}

\begin{figure}	
\epsfig{file=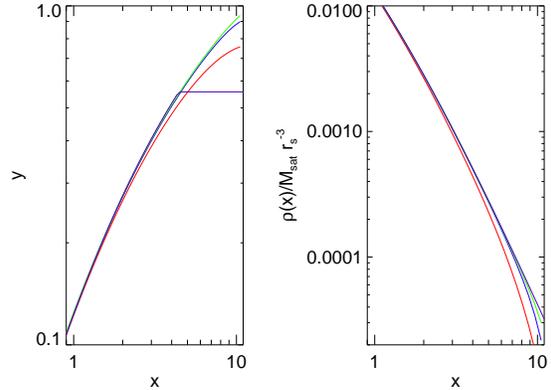,width=8cm}
\caption{The satellite's bound mass distribution (left panel) and mass density distribution (right panel) according to the model described in \S\ref{sec:phspmod} for different values of velocity anisotropy parameter ${\beta}=0.$ ({\it red} line), ${\beta}=0.6$ ({\it blue} line), and ${\beta}=0.9$ ({\it green} line). The purple line on the left panel corresponds to the predictions from the classical approach, on the right panel the initial---prior to stripping---satellite profile ({\it purple} line) is given for reference. The results are shown after one mass loss iterations. The tidal field strength is $f_{\rm tidal}=5.8\times 10^{-3}$. Results are shown for an NFW satellite with concentration parameter $c=11$.}
\end{figure}

\begin{figure}	
\epsfig{file=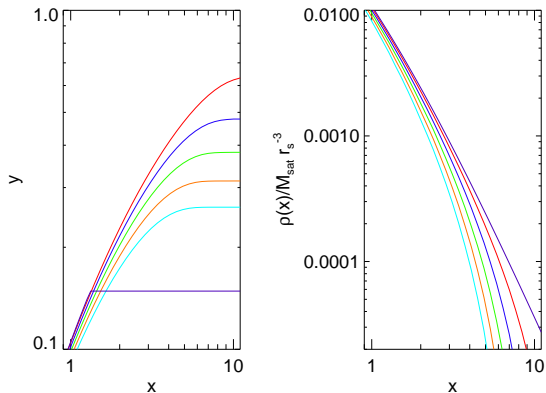,width=8cm}
\caption{The evolution of the satellite's bound mass distribution (left panel) and mass density distribution (right panel) according to the model described in \S\ref{sec:phspmod} for a concentration parameter $c=15$. The purple line on the left panel corresponds to the predictions from the classical approach. The results are shown after 1, 2, 3, 4 and 5 mass loss iterations. The tidal field strength is $f_{\rm tidal}=5.8\times 10^{-3}$. Results are shown for an NFW satellite with velocity anisotropy parameter ${\beta}=0$.}
\end{figure}

\begin{figure}	
\epsfig{file=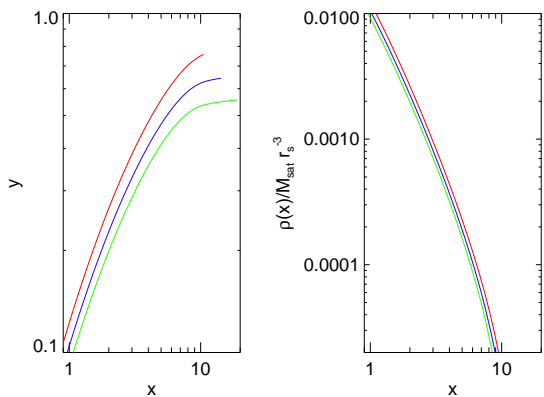,width=8cm}
\caption{The satellite's bound mass distribution (left panel) and mass density distribution (right panel) according to the model described in \S\ref{sec:phspmod} for different values of concentration parameter $c=11$ ({\it red} line), $c=15$ ({\it blue} line), and $c=20$ ({\it green} line). The purple line on the left panel corresponds to the predictions from the classical approach. Results are shown after one mass loss iterations. The tidal field strength is $f_{\rm tidal}=5.8\times 10^{-3}$. Results are shown for an NFW satellite with velocity anisotropy parameter ${\beta}=0$.}
\end{figure}

\subsubsection{Calculation of the velocity dispersion}

We find the velocity dispersion of the satellite by solving the Jeans equation:
\begin{equation}
{\frac{1}{\rho(x)}}
{\d \over \d x} \rho(x) \sigma_{\rm r}^2(x)
+2{\frac{{\beta}{\sigma_{\rm r}}^{2}(x)}{x}}=-{\frac{d{\Phi(x)}}{dx}},
\label{eq:jeans}
\end{equation}
where $\sigma_{\rm r}(x)$ is the radial velocity dispersion.
For the radial anisotropy (defined as $\beta=1-\sigma_\phi^2/\sigma_r^2$, where $\sigma_\phi(x)$ is the tangential velocity dispersion) we use a fixed value (i.e. independent of radius), ranging from 0 (isotropic motion) to
$\beta=1$ (radial motion). Integration is begun at three times the virial radius, which is sufficiently large to give an accurate
determination of the velocity dispersion throughout the satellite.

\subsubsection{Calculation of the timescale for mass loss}

We wish to estimate the time over which the mass will be lost from the satellite. We expect that any unbound particle will be lost in a time of order
the dynamical time. Defining the dynamical time as
\begin{equation}
t_{\rm dyn}(x) = {\pi}{\sqrt{\frac{r_{\rm s}^{3} x^3}{GM_{\rm s}y(x)}}},
\end{equation}
an approximate mass loss rate from the
satellite is given by
\begin{equation}
{\dot{M}_{\rm loss}}={{\int}_{0}^{x_{\rm max}}}4{\pi}r_{\rm s}^3 x^2 {\rho}_{\rm loss}(x)t_{\rm dyn}^{-1}(x)\d x,
\end{equation}
where $\rho_{\rm loss}(x)$ is the density distribution of mass lost through tidal stripping (i.e. the original density profile minus that after tidal mass loss). The characteristic timescale for mass loss is then
\begin{equation}
\langle t_{\rm dyn}\rangle = f_\tau M_{\rm loss}/{\dot{M}_{\rm loss}},
\end{equation}
where we have introduced a parameter $f_\tau$ to allow this timescale to be adjusted to match numerical results. We expect
$f_\tau\sim 1$.

\subsubsection{Continuous mass loss}

Using the model described above, we can compute the mass lost by the satellite due to tidal forces, and the resulting density
profile of the remaining bound material. This new density profile will give rise to a new potential $\Phi(r)$, weaker than that of
the original satellite. As such, if we apply our same calculation of mass loss to the remaining bound particles, using this new
potential, we expect that further mass loss will occur. This process can, in principle, be repeated \emph{ad infinitum}. Assuming
that the potential changes on a timescale of approximately $\langle t_{\rm dyn}\rangle$ we can use our model to compute mass loss
as a function of time. Essentially, we are breaking up the continuous process of mass loss into discrete intervals (each of order
the dynamical time). We compute the total mass loss over each interval, before updating the satellite density profile and
potential prior to computing mass loss for the next interval. We will refer to these intervals as ``mass loss iterations''.

\subsection{Limitations of the Model}

Our model, while a vast improvement over the ``classic'' calculation of tidal mass loss, has its own limitations:
\begin{enumerate}
\item In our model there is the implicit assumption that the mass loss takes place instantaneously after a time $\langle t_{\rm
dyn}\rangle$. Since mass loss is a continuous process our choice to
update the satellite's potential only after a time $\langle t_{\rm dyn}\rangle$ is an approximation. Our model could be modified to use smaller steps in time, for example stepping forward by 10\% of the mass loss timescale, while removing only 10\% of the mass predicted to have been lost. This would result in more frequent updates in the density and potential thereby improving this approximation.

\item In computing the energy gained by each particle, we evaluate the tidal perturbation by integrating around the orbit which
the particle would follow in the absence of any tidal field. In reality, the orbit will be stretched by the tidal field, which
will in turn alter the energy gain. These higher order contributions to the energy gain are ignored in this work, although
in principle they could be comparable to the first order energy gain.

\item After each mass loss iteration, we hold the density profile of the remaining particles fixed and compute a new velocity dispersion profile using eqn.~(\ref{eq:jeans}). This effectively changes the energies and angular momenta of the remaining bound particles. A correct calculation of the phase space distribution of the particles after each episode of mass loss is an important missing ingredient of our model. We defer detailed study of this issue to a future paper (see \S\ref{sec:discuss} for further discussion of this point).
\end{enumerate}

\subsection{Basic Results}

In this section will present the basic results of the model discused in \S\ref{sec:phspmod}. Unless otherwise stated, we will consider an NFW satellite with concentration parameter $c_{\rm NFW}=11$, isotropic orbits ($\beta=0$) in an $f_{\rm tidal}=5.8\times 10^{-3}$ tidal field.

Figure 3 shows the evolution of the mass density distribution of the satellite. The results are shown after 1 ({\it red} line), 2 ({\it blue} line), 3 ({\it green} line), 4 ({\it orange} line) and 5 ({\it cyan} line) mass loss iterations, and for different values of the tidal field: $f_{\rm tidal}=5.8\times10^{-4}$ (left panel), $5.8\times10^{-3}$ (middle panel) and $5.8\times10^{-2}$ (right panel). The initial---prior to stripping---satellite profile ({\it purple} line) is given for reference. As expected, stronger tidal fields result in greater mass loss from the satellite. For the strongest tidal field shown, significant mass loss occurs even at the scale radius ($x=1$) of the satellite.

Figure 4 shows the corresponding evolution of satellite's bound mass distribution. The results are shown after 1 ({\it red} line), 2 ({\it blue} line), 3 ({\it green} line), 4 ({\it orange} line) and 5 ({\it cyan} line) mass loss iterations, and for different values of the tidal field: $f_{\rm tidal}=5.8\times10^{-4}$ (left panel), $5.8\times10^{-3}$ (middle panel) and $5.8\times10^{-2}$ (right panel). In the middle and left plots the purple line corresponds to the predictions from the classical approach. This method predicts zero mass loss for the plot in the left panel as the tidal radius is larger than the virial radius of the satellite. Note that our model predicts that the satellite continues to lose mass in each successive mass loss iteration---there is no evidence that the mass is converging to some fixed value. This is despite the fact that the tidal field is constant and is due to the weakening of the satellite's potential after each mass loss iteration. The classic model for tidal mass loss predicts a fixed mass for the satellite after tidal stripping. In our model, the satellite falls below this mass after two mass loss iterations for (except for the weakest tidal field shown, in which case the classic model predicts no mass loss).

In Figs.~5 and 6 we explore the dependence of mass loss on the velocity anisotropy parameter ${\beta}$. More specifically, Fig.~5 shows the evolution of the satellite's bound mass distribution (left panel) and mass density distribution (right panel) for a velocity anisotropy parameter ${\beta}=0.6$. The purple line on the left panel corresponds to the predictions from the classical approach, on the right panel the initial---prior to stripping---satellite profile ({\it purple} line) is given for reference. The results are shown after 1 ({\it red} line), 2 ({\it blue} line), 3 ({\it green} line), 4 ({\it orange} line) and 5 ({\it cyan} line) mass loss iterations.

Figure 6 shows the evolution of the satellite's bound mass distribution (left panel) and mass density distribution (right panel) for different values of velocity anisotropy parameter ${\beta}=0$ ({\it red} line), ${\beta}=0.6$ ({\it blue} line), and ${\beta}=0.9$ ({\it green} line). The purple line on the left panel corresponds to the predictions from the classical approach, on the right panel the initial---prior to stripping---satellite profile ({\it purple} line) is given for reference. The results are shown after one mass loss iteration. From Figs.~5 and 6 it can be seen that increasingly radial orbits result in less mass loss.

We note that our current calculation do not accurately account for the ellipticity of orbits (see Appendix~\ref{app:AM}). As such, we must treat this result with some caution. A more detailed analysis is deferred to a future paper in which the shapes of orbits will be accounted for.

In Figs.~7 and 8 we explore the dependence of mass loss on the concentration parameter $c_{\rm NFW}$. Figure 7 shows the evolution of the satellite's bound mass distribution (left panel) and mass density distribution (right panel) for a concentration parameter $c_{\rm NFW}=15$. The purple line on the left panel corresponds to the predictions from the classical approach. The results are shown after 1 ({\it red} line), 2 ({\it blue} line), 3 ({\it green} line), 4 ({\it orange} line) and 5 ({\it cyan} line) mass loss iterations.

Figure 8 shows the evolution of the satellite's bound mass distribution (left panel) and mass density distribution (right panel) for different values of concentration parameter: $c_{\rm NFW}=11$ ({\it red} line), $c_{\rm NFW}=15$ ({\it blue} line) and $c_{\rm NFW}=20$ ({\it green} line). The purple line on the left panel corresponds to the predictions from the classical approach. The results are shown after one mass loss iteration. From Figs.~7 and 8 it can be seen that more concentrated halos experience greater mass loss. Note, however, that we keep $f_{\rm tidal}$ fixed for each satellite considered here and use the NFW scale length as our unit of length, $r_{\rm s}$. As a result, halos with higher values of $c_{\rm NFW}$ actually have a smaller fraction of their total mass within $x=1$ and are more extended (i.e. their virial radii lie at larger values of $x$) making them more susceptible to mass loss. This does not necessarily reflect the scalings expected for cosmological halos for which the concentration and mass are correlated \cite{bullock01}.

\subsection{Comparison to Previous Work}
Here we will compare our method with the results presented in \cite{hayashi03}.
In particular, they found that one can get the satelite's density profile after tidal stripping starting with an NFW profile and taking into account two kind of modifications: a lowering of the central density $f$ and the introduction of a tidal radius $r_{te}$ of the outer profile. Hayashi et al. provide fitting curves both for the tidal radius $r_{te}$ and the lowering of the central density factor $f$ as a function of the satellite bound mass, so we are able to see if our results agree or not.

We choose as a comparison variable the tidal radius $r_{te}$ since it is a better defined quantity. Since  in our model we do not have a sharp cut-off radius we define $r_{te}$ as the radius at which the density drops to $50\%$ of the original NFW value. Figure~9 shows the tidal radius $r_{te}$ as a function of the satellite bound mass. The solid line corrsponds to the fitting curve provided by Hayashi et al., the coloured points are results from our model that correspond to tidal fields of different strength(red points, $f_{\rm tidal}=5.8\times 10^{-2}$, cyan points, $f_{\rm tidal}=5.8\times 10^{-3}$, and purple points, $f_{\rm tidal}=5.8\times 10^{-4}$). Our results, especially for the case of strong tidal filds are in good agreement with Hayashi et al. fitting curve.

\begin{figure}	
\epsfig{file=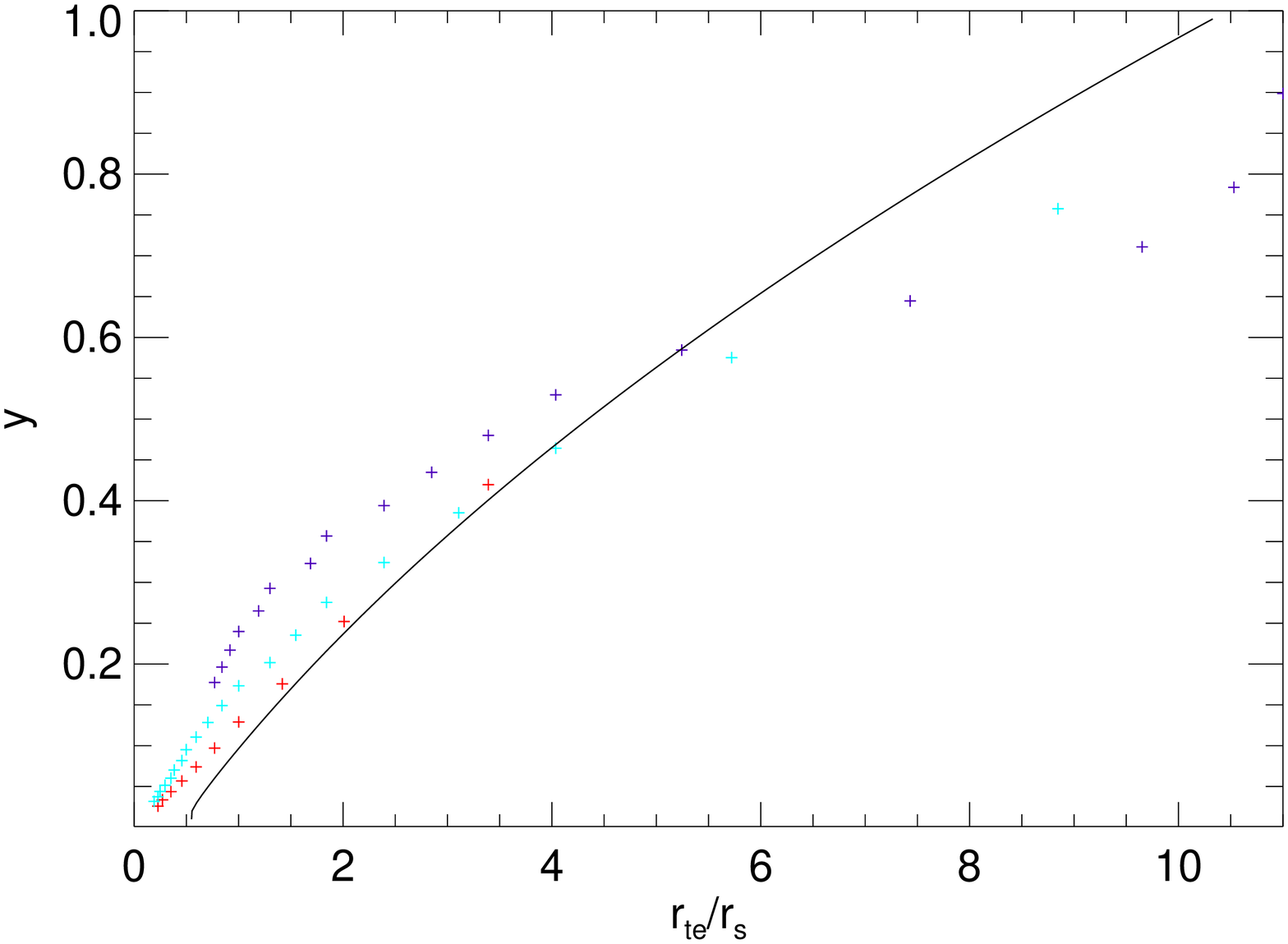,width=8cm}
\caption{The tidal radius $r_{te}$ as a function of the satellite bound mass. The solid line corrsponds to the fitting curve provided by Hayashi et al., the coloured points are results from our model that correspond to tidal fields of different strength(red points, $f_{\rm tidal}=5.8\times 10^{-2}$, cyan points, $f_{\rm tidal}=5.8\times 10^{-3}$, and purple points, $f_{\rm tidal}=5.8\times 10^{-4}$).}
\end{figure}

\section{Comparison with N-Body Simulations}
\label{sec:nbody}

In order to test the ability of our model to quantitatively describe mass loss from dark matter haloes we have run several
N-body simulations. Simulations were performed using {\sc Gadget2} \cite{gadget2}, modified to include a point mass at the centre
of the coordinate system. To study mass loss, an N-body representation of an NFW dark matter halo with a concentration parameter
of $11$ was placed in a circular orbit around this point mass. A softening length equal to 14\% of the scale length was chosen
(this is the optimal softening length suggested by \pcite{poweretal}). To ensure accurate evolution\footnote{The reader is
referred to the {\sc gadget2} user guide for definitions of numerical parameters.}, we set the maximum timestep to be 13\% of the
dynamical time at the scale radius (corresponding to 1\% of the dynamical time at the virial radius) and set {\tt
ErrTolIntAcc}$=0.025$. A force accuracy of {\tt ErrTolForceAcc}$=0.005$ was chosen.

We have performed numerous tests to ensure that our simulations are not affected by the choice of numerical parameters in {\sc
Gadget2}. Key numerical parameters such as the softening length, maximum timestep, integration accuracy and force accuracy were
reduced by factors of two and the calculations repeated. In no case do we see any significant change in the evolution of the
satellite mass or density profiles. As such, the mass loss seen in these calculations is due entirely to the tidal field from the
point mass. Furthermore, with these parameters, the orbital radius of the centre of mass of the satellite varies by only 0.06\%
throughout the duration of our simulation. This corresponds to a 0.18\% variation in $f_{\rm tidal}$.

The mass and orbital radius combination of the circular orbit were chosen to produce a desired $f_{\rm tidal}$. The satellite was
allowed to orbit for a period corresponding to over 85 dynamical times at the scale radius (or, equivalently, almost 7 dynamical
times at the virial radius). At each timestep of the calculation we performed a binding energy analysis on all satellite particles
which were bound at the end of the preceding step to check whether any of these had become unbound. (We define ``bound'' here to mean that the particle's energy in a frame coinciding instantaneously with the centre of mass of the satellite, and including gravitational energy only from other bound particles, is negative.) The remaining bound particles
were used to compute a density profile and total mass for the satellite.

Satellites were set up using $10^5$ particles sampled from the phase space density described in \S\ref{sec:phspmod}, and with
particles out to three times the virial radius.

To test for numerical convergence, we began by running our calculations using smaller numbers of particles. The particle number was gradually increased until converged results were obtained. Using $3\times 10^4$ particles, for example, we find that the remaining bound mass of the satellite differs
from that in the $10^5$ particle simulation by less than 2\% at all times. As such, $10^5$ particles are enough to ensure that particle number does not significantly affect
our results.

\begin{figure}
\psfig{file=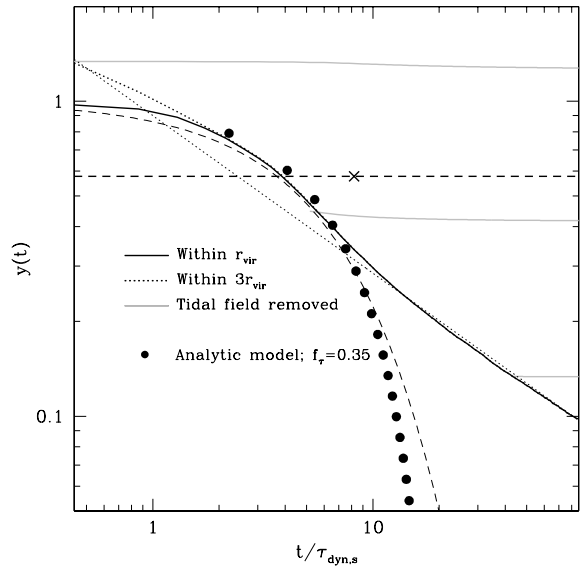,width=80mm}
\caption{The mass, $y(t)$ (defined as the remaining bound mass divided by the initial bound mass of the satellite), of a $c=11$ satellite in an $f_{\rm tidal}=5.8\times 10^{-3}$ tidal field as a function of time (in units of the
dynamical time at the satellite scale radius). The solid black line shows the remaining bound mass of the satellite within the
virial radius as determined by N-body simulation. The dotted black line indicates the total bound mass of the satellite
(i.e. including particles beyond the virial radius). Grey lines indicate how the mass of the satellite evolves when the tidal
field is removed at various points in time. The horizontal dashed line indicates the mass within the classic tidal
radius, while the cross on this line indicates the mass loss timescale (for $f_{\tau_{\rm dyn}}=1$) for the classic calculation. Points indicate the mass predicted by the model described in this work after 1--17 iterations. Thin dashed and dotted lines are shown to illustrate the two distinct regimes of mass loss seen in the N-body simulation---an initial rapid regime in which mass declines almost exponentially with time (dashed line) followed by a slower regime in which mass declines as a power-law in time (dotted line).}
\label{fig:Nbody}
\end{figure}

Figure~\ref{fig:Nbody} shows the mass of our satellite in an $f_{\rm tidal}=5.8\times 10^{-3}$ tidal field as a function of
time. The solid black line indicates the remaining bound mass of the satellite within the virial radius. For reference, the dotted
black line indicates the total bound mass of the satellite (i.e. including particles beyond the virial radius)---unsurprisingly
mass beyond the virial radius is quickly stripped away. At several times during our simulation we have extracted the particles
which remain bound at that time and evolved them in isolation (i.e. without the presence of unbound particles \emph{and} without
any tidal field applied). Results from three such calculations are shown as solid grey lines in Fig.~\ref{fig:Nbody}. With the
tidal field removed, the mass of the satellite quickly reaches a constant value\footnote{Note that the mass does decline slightly even after the tidal field is switched off. This transient mass loss is due to the fact that the satellite must relax to a new configuration after the tidal potential is removed, leading to some particles gaining energy and becoming unbound.}, providing further evidence that numerical effects
do not contribute to the mass loss from this satellite, which must instead be entirely due to the applied tidal field. The fact
that the satellite mass remains close to that at the time at which the host potential was removed validates our criterion for
deciding whether or not a particle should be considered to be bound to the satellite or not.

Inspection of the N-body mass loss curve indicates that there are two distinct regimes of mass loss: an initial rapid regime in which mass declines almost exponentially with time (illustrated by the thin dashed line) followed by a slower regime in which mass declines as a power-law in time (illustrated by the thin dotted line).

The results of analytical calculations of mass loss are also shown in Fig.~\ref{fig:Nbody}. The horizontal dashed line indicates
the mass of the satellite after tidal mass loss according to the classical model. The cross on this line indicates the mean
dynamical time for mass loss in this model for $f_{\tau_{\rm dyn}}=1$ (note that we could always force this cross to lie on the
N-body line in the classical model by suitable choice of $f_{\tau_{\rm dyn}}$). The limitations of the classical model are clearly
seen---the satellite loses mass continuously throughout the calculation, while the classical model predicts a fixed mass.

Circles indicate the prediction from the mass loss model described in this work. We show results for 17 mass loss iterations and
chose $f_{\tau_{\rm dyn}}=0.35$. As expected, $f_{\tau_{\rm dyn}}$ is of order unity. Note that our model is able to match the
rate of mass loss quite well for about five iterations, before beginning to significantly overestimate the rate of mass loss. This discrepancy will be discussed further in \S\ref{sec:discuss}.

\begin{figure}
\psfig{file=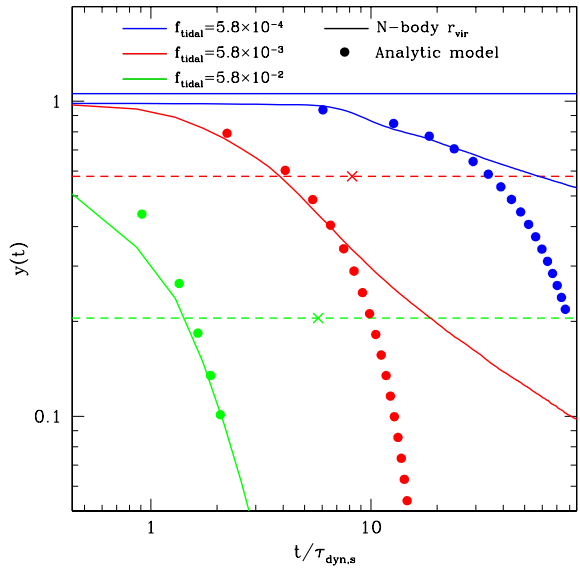,width=80mm}
\caption{The mass of a $c=11$ satellite in tidal fields of varying strengths as a function of time (in units of the dynamical time
at the satellite scale radius). Solid lines show the remaining bound mass of the satellite within the virial radius as determined
by N-body simulation. The horizontal dashed lines indicate the mass within the classic tidal radius. Points indicate the mass
predicted by the model described in this work after 1--5 iterations.}
\label{fig:NbodyVarM}
\end{figure}

Figure~\ref{fig:NbodyVarM} shows mass loss as a function of time for the same satellite in tidal fields of three different
strengths. Coloured lines indicate the bound mass of the N-body satellite, while coloured points indicate the results from the
model described in this work. Dashed lines with crosses indicate the results of the classical model as in Fig.~\ref{fig:Nbody}. To
obtain a good fit to the mass loss rates we are forced to adopt different values of $f_\tau$ depending on the strength of the
tidal field. The results show have $f_\tau=1.4$, 0.35 and 0.15 for tidal fields of $f_{\rm tidal}=5.8\times 10^{-4}$, $5.8\times
10^{-3}$ and $5.8\times 10^{-2}$ respectively. Once again we see that our model describes the mass loss quite well for several
iterations before overestimating the mass loss rate at late times. (This overestimation is not seen for the $f_{\rm tidal}=5.8\times 10^{-2}$ calculation as the satellite loses its mass so rapidly that the second, slow mass loss regime is never seen.)

\begin{figure}
\psfig{file=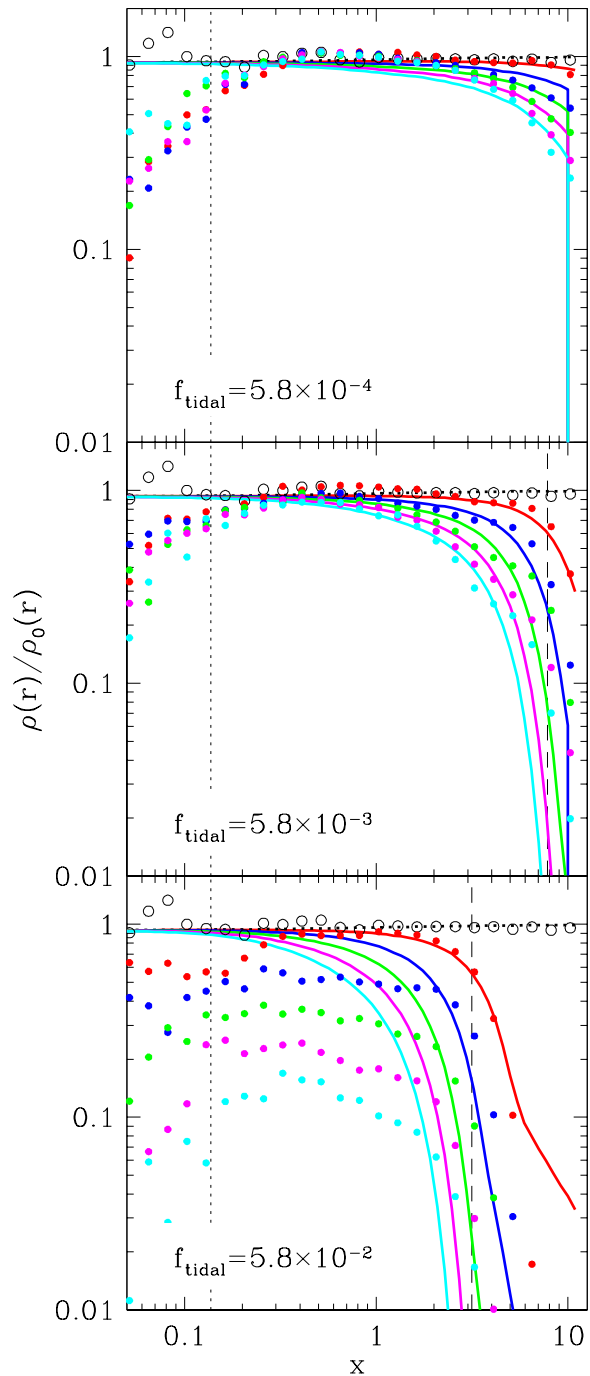,width=80mm,bbllx=0mm,bblly=15mm,bburx=105mm,bbury=265mm,clip=}
\caption{The density profile of the satellite galaxy is shown relative to the original profile and as a function of radius (in
units of the satellite scale radius). The three panels show results for three different values of $f_{\rm tidal}$: $5.8\times
10^{-4}$ (upper panel), $5.8\times 10^{-3}$ (centre panel), $5.8\times 10^{-2}$ (lower panel). Lines show predictions from the
calculations described in this work, while circles indicate results from the N-body simulations. Results are shown after 0--5
iterations as indicated by colour (black, red, blue, green, magenta, cyan). Vertical dotted black lines indicate the softening
length in the N-body simulations. Vertical dashed black lines indicate the classic tidal radius.}
\label{fig:NbodyDensity}
\end{figure}

In addition to the total bound mass of the satellite, our model is able to predict the radial density profile of that mass. In
Fig.~\ref{fig:NbodyDensity} we show as coloured lines the density profile (normalized to the original density profile) of the same
satellite in three different tidal fields after 1--5 iterations. Coloured circles show the density profile from the N-body
simulation at the corresponding times. Vertical dotted lines indicate the softening length in our calculations. As expected, in
the N-body simulations the density at radii less than a few times the softening length drops quickly. The vertical dashed line
indicates the tidal radius in the classic model (this line cannot be seen in the upper panel as the classical model predicts a
tidal radius beyond the virial radius for this tidal field strength).

For the weakest tidal field (upper panel of Fig.~\ref{fig:NbodyDensity}) the analytic model predicts the radial density profile
seen in the N-body simulation reasonably accurately for the first iteration (red line). After this, our
model underpredicts the mass loss rate and, as such, overpredicts the resulting density profile. As the tidal field is increased
the model fairs worse. For the strongest field considered we can see that the N-body satellite quickly develops a constant density
core, the density of which decreases with time. Our model does not reproduce this behaviour. This may reflect the fact that our
model applies only first order perturbations to the particle orbits. As $f_{\rm tidal}$ is increased the orbits of particles which
remain bound to the satellite become ever more perturbed due to the tidal force. A measure of this perturbation can be constructed
by averaging $-\Delta E/E$ (see eqn.~\ref{eq:deltaE} for the definition of $\Delta E$) over all particles which remain bound, which measures the fractional change in particle orbital
energies. This quantity, after one iteration, is 0.08, 0.21 and 0.30 for $f_{\rm tidal}=5.8\times 10^{-4}$, $5.8\times 10^{-3}$
and $5.8\times 10^{-2}$ respectively. Thus, for the strongest field considered this factor is becoming quite large. Furthermore,
for an NFW potential, the gravitational potential varies only very slowly with radius at radii less than the scale radius. Thus, a
relatively small perturbation in the energies of particles at these radii can lead to a large perturbation in their apocentric
distance. As a result, the second order correction to the work done by the tidal field can be large.

\section{Discussion and conclusions}
\label{sec:discuss}

We have presented a model of tidal mass loss from collisionless systems with arbitrary phase space distributions. Our calculation has many advantages over the classic model of tidal mass loss (in which the density profile is truncated beyond a tidal radius determined through force balance arguments). In particular, our calculation takes into account the orbital structure of the system, permitting the effects of anisotropic orbits on the degree of mass loss to be investigated. Furthermore, we are able to estimate the density profile of the material remaining after mass loss has occurred.

A key prediction from this model is that mass loss will be continuous even in a static field---the bound mass of the system shows no sign of converging to a fixed value. This behaviour is also seen in the N-body simulations which we have carried out. In particular, the N-body simulations show evidence for two distinct regimes of mass loss: an initial rapid phase in which the mass declines exponentially with time followed by a slower phase during which the mass declines as a power-law in time.

The rate of mass loss predicted by our model is comparable to that seen in the N-body simulations during the initial, rapid mass loss phase, although we find that the parameter $f_\tau$ (which scales the rate of mass loss in our model) must vary with the tidal field strength in order to match N-body results. We intend to explore and characterize this systematic variation of $f_\tau$ in a future paper. This continuous mass loss is a consequence of the fact that mass loss weakens the gravitational potential of the system thereby allowing particles to become unbound that could not have escaped from the original potential. At larger times, while the N-body simulations transition to a slower mass loss regime our model continues to show a rapid, exponential mass loss. The origin of the slow mass loss regime in the N-body simulations and the failure of our model to reproduce this is not understood at present, but is the focus of ongoing study. Nevertheless, our model is able to describe the rate of mass loss quite well over almost 10 satellite dynamical times.

The model presented in this work predicts small amounts of mass loss even when the tidal field is sufficiently weak that the classic model predicts zero mass loss (due to the tidal radius lying beyond the outer radius of the system). As this mass loss occurs continuously it can eventually lead to a large reduction in the mass of the system (e.g. in the left-hand panel of Fig.~\ref{fig:varftidal}, with $f_{\rm tidal}=5.8\times 10^{-4}$, the mass has been reduced to almost 60\% of its original value after five mass loss iterations). This could have important consequences for the distribution and survival of low-mass satellite systems in dark matter haloes.

In Fig.~\ref{fig:NbodyDensity} we see that, for the strongest tidal field, our model fails to predict the reduction in density in
the centre of the satellite. This may explain the rather low value of $f_\tau$ we require to fit the mass loss rate in this
case. If our model correctly predicted the mass loss from the centre of the halo it would naturally predict a lower $\langle \tau
\rangle$. As noted in \S\ref{sec:nbody} this may be due to the fact that our estimate of the tidal torque exerted on each particle is correct only to first order---an improved calculation accounting for the distortion of particle orbits by tidal torques should result in greater mass loss, and is expected to be particularly important at small radii. We defer study of this issue to a future paper.

We have demonstrated that our model is able to consider the effects of orbital anisotropy and profile shape on the degree of mass loss. A full study of the effects of anisotropy must await a full treatment of elliptical orbits, to be presented in a future paper. 

The major shortcoming of the present model is the lack of a reasonable method to compute the phase space density of particles after each mass loss iteration. Currently, we simply assume that the density profile of the remaining bound material remains fixed (i.e. the radial distribution of bound particles after mass loss is identical to the radial distribution of the same particles prior to mass loss). We then simply re-solve Jeans equation to find a velocity dispersion which results in an equilibrium system built from these particles. 

A possible improvement upon this may be to treat the potential of the satellite as varying only slowly with time. Changes in particle orbits can then be considered in terms of adiabatic invariants and a new phase space density constructed which explicitly conserves these invariants for bound particles after mass loss has occurred. It should be noted, however, that mass loss is assumed to occur on a time comparable to the local dynamical time in the satellite (an assumption confirmed by the N-body simulations carried out here). As such, the changes in the potential occur on a timescale comparable to the orbital times of particles, making the adiabatic approximation a poor one. The success of such an approach therefore remains to be determined.

In a future paper, we will develop this model further. Along with the improvements mentioned above, we intend to examine more carefully the dependence of the energy gain through tidal torques on orbital eccentricity and the effects of using smaller time intervals for each mass loss iteration\footnote{Preliminary investigations suggest that using smaller time intervals does not change our results}. We will also explore the application of our model to two-component (i.e. dark matter + stellar) systems and present fitting formula for mass loss as a function of tidal field strength and time. A further important consideration will be to apply our model to situations in which the tidal field varies with time.

The model of tidal mass loss presented here represents part of an ongoing effort to improve the accuracy and predictive power of analytic models of satellite orbits and evolution. Detailed modelling of this sort is required in order to understand the substructure of dark matter halos and the evolution of the galaxies within them.

\section*{Acknowledgements}

AJB acknowledges support from a Royal Society University Research Fellowship. We acknowledge valuable discussions with Scott Kay,
Joe Silk, John Magorrian and James Binney.

\appendix

\section{Angular Momentum Perturbations}
\label{app:AM}

This Appendix details our calculation of the effects of tidal forces on particles orbitting in a satellite based upon perturbations to a particle's angular momentum.

\subsection{Tidal Torques}

We consider a satellite galaxy placed in a circular orbit within a spherical host potential. Effects of dynamical friction are
neglected. We consider a frame which rotates at the same rate as the satellite orbits.

The gravitational force (per unit mass) at a position $\mathbf{r^\prime}$ from the centre of the host potential is
\begin{equation}
{\bf F}_{\rm g}({\bf d}) = - {\G M^\prime(d) \over d^2} {\bf \hat{d}},
\end{equation}
where $M^\prime(d)$ is the mass of the host contained within radius $d$ and a hat indicates a unit vector.

In addition, in this rotating frame, the particle experiences a fictitious force of
\begin{equation}
{\bf F}_\Omega = \Omega^2 {\bf d} - 2 \mathbf{\Omega} \wedge {\bf v},
\end{equation}
where $\mathbf{\Omega}$ is a vector with magnitude equal to the angular frequency of the satellite orbit, and normal to the
orbital plane, ${\bf v}$ is the velocity of the particle. For the circular orbit considered,
\begin{equation}
\Omega = \sqrt{\G M^\prime(d)/d^3}.
\end{equation}
The centrifugal force here will act to cancel the mean gravitational force on the satellite since 
\begin{equation}
\Omega^2 d =  {\G M^\prime(d) \over d^2}.
\end{equation}
At any point in the satellite, we are therefore left with a tidal force (i.e. the gravitational force minus this mean
gravitational force) which has a quadrupole form, the Coriolis force and an internal force due to the mass of the satellite
itself. This latter will be ignored since it exerts no torque on the particle.

We consider a particle orbitting within the satellite. Let the satellite centre be at ${\bf r}^\prime$, and the particle be at
${\bf d}$. The orbital position of the particle relative to the satellite centre is ${\bf r}={\bf d} - {\bf r}^\prime$. We are
interested in the angular momentum of the particle around the satellite centre and so wish to compute the torque exerted on the
particle relative to this centre:
\begin{eqnarray}
{\bf T} & = & {\bf r} \wedge [{\bf F}_{\rm g}({\bf d}) + {\bf F}_\Omega] \nonumber \\
        & = & {\bf r} \wedge \left[  - {\G M^\prime(d) \over d^2} {\bf \hat{d}} +  {\G M^\prime(r^\prime) \over r^{\prime 2}} {\bf \hat{r}}^\prime - 2 \mathbf{\Omega} \wedge {\bf v} \right].
\end{eqnarray}
Considering just the tidal part of the above, the magnitude of the tidal torque is
\begin{equation}
T_{\rm tidal} = r \left[ {\G M^\prime(d) \over d^2} \sin \beta - {\G M^\prime(r^\prime) \over r^{\prime 2}}\sin\theta  \right],
\end{equation}
where $\theta$ is the angle\footnote{Note that ${\bf r}\wedge{\bf r}^\prime$ and ${\bf r}\wedge{\bf d}$ point in the same
direction.} between ${\bf r}$ and ${\bf r}^\prime$ and $\beta$ is the angle between ${\bf r}$ and ${\bf d}$. Defining
\begin{equation}
c_{\rm tidal} = {r^{\prime 3} \over \G M^\prime(r^\prime)} {\d \over \d r^\prime} \left( {\G M^\prime(r^\prime) \over r^{\prime 2}} \right)
\end{equation}
(e.g. $c_{\rm tidal}=-2$ for a point mass), we can write
\begin{equation}
{\G M^\prime(d) \over d^2} \approx {\G M^\prime(r^\prime) \over r^{\prime,2}} \left[1+c_{\rm tidal}{d-r^\prime\over r^\prime}\right]
\end{equation}
which is the usual linear approximation for the tidal field, valid providing $r\ll r^\prime$. Therefore,
\begin{equation}
T_{\rm tidal} = r {\G M^\prime(r^\prime) \over r^{\prime,2}} \left[
\left(1+c_{\rm tidal}{d-r^\prime\over r^\prime}\right)
 \sin \beta - \sin\theta  \right].
\end{equation}
If $\alpha$ is the angle between ${\bf r}^\prime$ and ${\bf d}$, then
\begin{equation}
{\sin \alpha \over r} = {\sin \theta \over d}
\end{equation}
and $\beta = \pi-\theta-\alpha$. We also have
\begin{equation}
d^2 = r^{\prime,2}+r^2-2r^\prime r\cos\theta.
\end{equation}
Combining these results, we find
\begin{eqnarray}
T_{\rm tidal} & = & r {\G M^\prime(r^\prime) \over r^{\prime,2}} \left[
\left(1+c_{\rm tidal}{d-r^\prime\over r^\prime}\right)
 \right. \nonumber \\
 & & \left. \times \sin \left( \theta + \sin^{-1} \left\{{r\over d}\sin\theta\right\} \right) - \sin\theta  \right].
\end{eqnarray}
Substituting for $d$ and expanding as a series in $r/r^\prime$ we find
\begin{eqnarray}
T_{\rm tidal} & = & r {\G M^\prime(r^\prime) \over r^{\prime,2}} \left[
 (1-c_{\rm tidal})\cos\theta\sin\theta {r \over r^\prime} \right. \nonumber \\
 & &  + {1 \over 2} (3\cos^2\theta -1)(1-c_{\rm tidal}) \sin\theta {r^2 \over r^{\prime 2}} \nonumber \\
 & & \left. + \mathcal{O}\left( {r^3 \over r^{\prime 3}}\right)
\right] .
\end{eqnarray}
The first term in this expression shows the quadrupole nature of the tidal torque. Note that the tidal torque depends on $c_{\rm
tidal}$ as expected, but there is an additional contribution (i.e. we have $1-c_{\rm tidal}$ rather than just $-c_{\rm tidal}$)
which arises from the fact that, even if there is no gradient in the gravitational force as a function of distance from the host
centre, there is still a difference in the \emph{vector} forces acting at the satellite centre and at the position of the
particle, which acts as a tidal field. We will ignore higher order terms from now on.

We now wish to find the change in the angular momentum of the particle as it moves around its orbit in the satellite galaxy. Note
that the tidal torque always acts normal to the plane containing ${\bf r}^\prime$ and ${\bf r}$. The change in angular momentum
around the orbit is given simply by
\begin{equation}
\Delta {\bf j} = \int {\bf T}_{\rm tidal} \d t.
\end{equation}
Writing $\d t = \d \chi / \dot{\chi}$ where $\chi$ is an angle measured around the orbit, and using the fact that $\dot{\chi} =
j_0/r(\chi)^2$, where $j_0=rv\sin\phi$ is the unperturbed angular momentum of the particle, we have
\begin{equation}
\Delta {\bf j} = \int {\bf T}_{\rm tidal} {r^2(\chi) \over j_0} \d \chi.
\end{equation}
For an orbit in a plane whose normal makes an angle $\xi$ with ${\bf r}^\prime$, we have
\begin{equation}
\cos\theta = \cos\chi \sin\xi,
\end{equation}
where we have chosen $\chi$ to coincide with $\theta$ for the case $\xi=\pi/2$. Thus
\begin{eqnarray}
\Delta {\bf j} & = & {\G M^\prime(r^\prime) \over j_0 r^{\prime 3}} (1-c_{\rm tidal}) \nonumber \\
 & & \int
{\bf \hat{T}} \cos\chi \sin\xi (1- \cos^2\chi \sin^2\xi)^{1/2}
 r^4(\chi) \d \chi.
\end{eqnarray}
To assess the vector change in ${\bf j}_0$ it is useful to consider three orthogonal unit vectors defined by ${\bf \hat{r}}^\prime$,
$({\bf \hat{r}}^\prime\wedge {\bf \hat{j}}_0)/\sin\xi$ and $[{\bf \hat{r}}^\prime \wedge ({\bf \hat{r}}^\prime\wedge {\bf
\hat{j}}_0)]/\sin\xi$. The angular momentum vector ${\bf j}_0$ has projections onto these three vectors of
\begin{equation}
{\bf j}_0 = \left( \begin{array}{c} j_0\cos\xi \\ 0 \\ j_0\sin\xi \end{array} \right) .
\end{equation}
The tidal torque unit vector has projections
\begin{equation}
{\bf \hat{T}} = \hbox{sign}(\sin\theta\cos\theta)\left( \begin{array}{c} 0 \\ |\cos\xi\cos\theta| \\ (1-\cos^2\xi\cos^2\theta)^{1/2} \end{array} \right) ,
\end{equation}
where
\begin{equation}
\hbox{sign}(x) = \left\{ \begin{array}{ll} -1 & \hbox{if } x < 0 \\ 0 & \hbox{if } x = 0 \\ 1 & \hbox{if } x > 0 \end{array} \right. ,
\end{equation}
and we have made explicit the quadrupole form of the torque. Therefore, the change in the angular momentum is
\begin{eqnarray}
\Delta {\bf j} & = & {\G M^\prime(r^\prime) \over j_0 r^{\prime 3}} (1-c_{\rm tidal}) \int r^4(\chi) \hbox{sign}(\sin\theta\cos\theta) \nonumber \\
 & & \times \cos\chi \sin\xi (1- \cos^2\chi \sin^2\xi)^{1/2} \nonumber \\
 & &  \times \left( \begin{array}{c}
0 \\
|\cos\chi \cos\xi \sin\xi| \\
|(1-\cos^2\xi\cos^2\chi \sin^2\xi)^{1/2}|
\end{array} \right) \d \chi .
\end{eqnarray}
Even for a circular orbit (i.e. constant $r$) the integrals are not analytically tractable in the general case. For $\xi=\pi/2$,
we find the change in angular momentum between $\chi_1$ and $\chi_2$ to be
\begin{equation}
\Delta {\bf j} = r^4 {\G M^\prime(r^\prime) \over j_0 r^{\prime 3}} (1-c_{\rm tidal}) \left( \begin{array}{c}
0 \\
0 \\
1/2 [\cos^2\chi]_{\chi_1}^{\chi_2}
\end{array} \right),
\end{equation}
and so, the maximum variation in $j$ is given by (e.g for $\chi_1=0$ and $\chi_2=\pi/2$)
\begin{equation}
j = j_0 \pm r^4 {\G M^\prime(r^\prime) \over 2 j_0 r^{\prime 3}} (1-c_{\rm tidal}).
\end{equation}

\begin{figure}
\epsfig{file=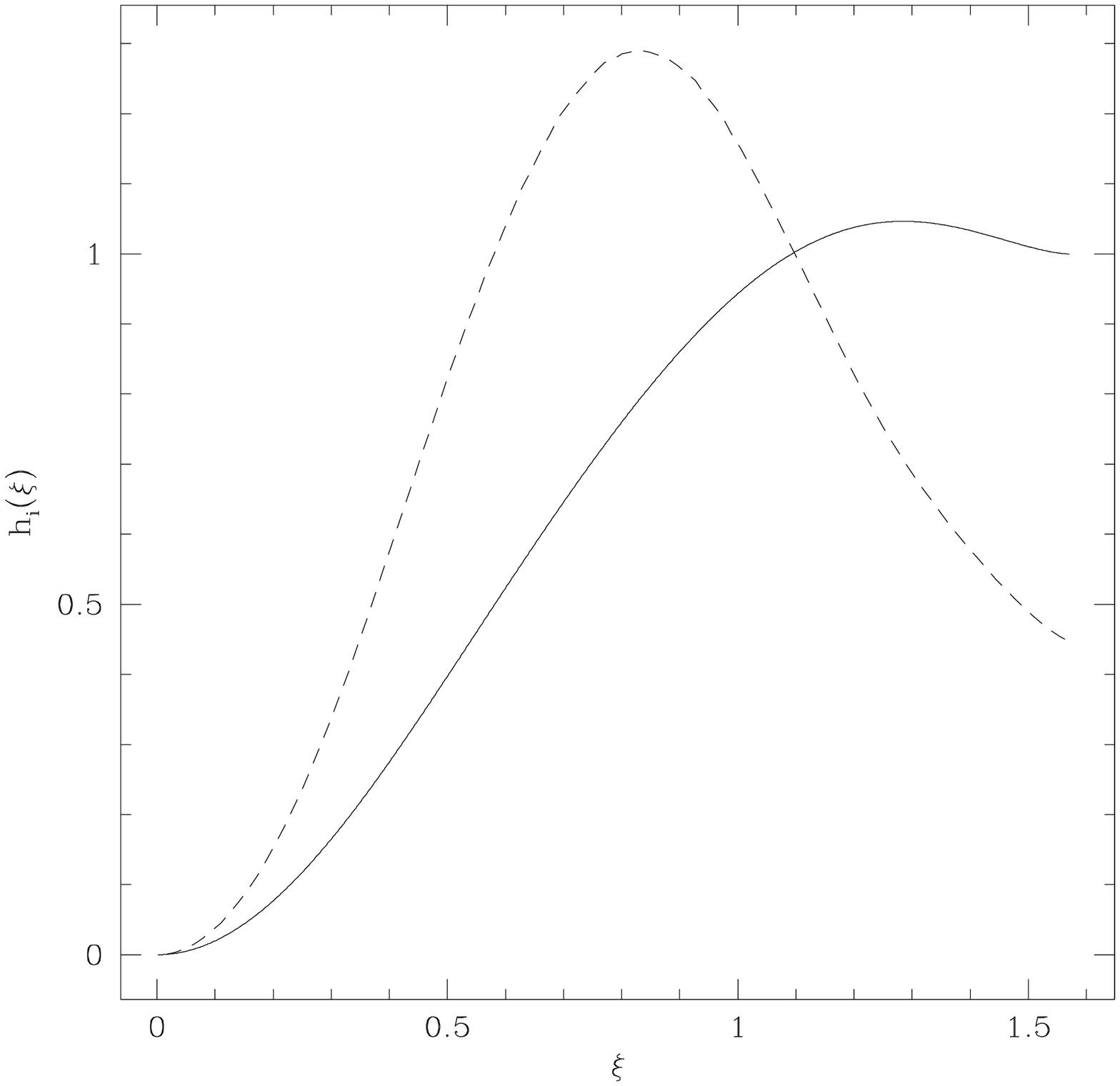,width=80mm}
\caption{The function $h_1(\xi)$ for the cases where Coriolis torques are not included is shown by the solid line. Results for the
case where Coriolis forces are included (and are averaged over $\gamma$) are shown by the dashed line.}
\label{fig:hxi}
\end{figure}

In general we can write
\begin{eqnarray}
\Delta {\bf j} & = & j_0 g_{\rm tidal} \int \left({r(\chi) \over r^\prime}\right)^4 \hbox{sign}(\sin\theta\cos\theta) \nonumber \\
 & & \times \cos\chi \sin\xi (1- \cos^2\chi \sin^2\xi)^{1/2} \nonumber \\
& & \times \left( \begin{array}{c}
0 \\
|\cos\chi \cos\xi \sin\xi|  \\
|(1-\cos^2\xi\cos^2\chi \sin^2\xi)^{1/2}|
\end{array} \right) \d \chi .
\end{eqnarray}
where the dimensionless number $g_{\rm tidal} = (\G M^\prime(r^\prime) r^\prime / j_0^2) (1-c_{\rm tidal})$. Then, the perturbed
angular momentum at $\chi$ is
\begin{equation}
{\bf j}(\chi) = {\bf j}_0 + \Delta {\bf j}(\chi).
\end{equation}
Considering the kinetic energy (per unit mass) of the particle in a frame moving with the satellite centre,
\begin{equation}
{1 \over 2} v^2(\chi) = {1 \over 2} {j^2(\chi) \over r^2(\chi)},
\end{equation}
we find
\begin{equation}
{1 \over 2} v^2(\chi) = {1 \over 2} v_0^2(\chi) \left[ 1 + 2 {\Delta {\bf j}\cdot {\bf j}_0 \over j_0^2} + {\Delta j^2 \over j_0^2}\right],
\end{equation}
and so the gain in energy is
\begin{equation}
\Delta E = {1 \over 2} {j_0^2 \over r^2(\chi)} \left[  2 {\Delta {\bf j}(\chi)\cdot {\bf j}_0 \over j_0^2} + {\Delta j^2(\chi) \over j_0^2}\right].
\end{equation}
Simplifying to circular orbits
\begin{equation}
\Delta E = {1 \over 2} {j_0^2 \over r^2} \left[  2 {\Delta {\bf j}(\chi)\cdot {\bf j}_0 \over j_0^2} + {\Delta j^2(\chi) \over j_0^2}\right].
\end{equation}
Integrating from $\chi=0$ to $\chi=\pi/2$ gives the maximum change in kinetic energy (as this maximizes both terms in the
above). Defining,
\begin{eqnarray}
h_1(\xi) & = & 2 \Delta {\bf j}(0,\pi/2)\cdot {\bf j}_0 / j_0^2 / g_{\rm tidal} /(r/r^\prime)^4\\
h_2(\xi) & = & \Delta j^2(0,\pi/2)/j_0^2/g^2_{\rm tidal}/(r/r^\prime)^8,
\end{eqnarray}
numerical integration leads to the forms shown in Fig.~\ref{fig:hxi}.
Writing out the energy expression in detail we find
\begin{eqnarray}
\Delta E & = & {1 \over 2} h_1(\xi) (1-c_{\rm tidal}) {\G M^\prime(r^\prime) \over r^\prime} \left( {r \over r^\prime} \right)^2 \nonumber \\
 & & + {1 \over 2} h_2(\xi) (1-c_{\rm tidal})^2 \left( {\G M^\prime(r^\prime) \over r^\prime}\right)^2 {r^{\prime 2} \over j_0^2} \left( r \over r^\prime \right)^6 
\end{eqnarray}
or
\begin{eqnarray}
\Delta E & = & {v^{\prime 2} \over 2} \left( {r \over r^\prime} \right)^2 \left[ h_1(\xi) (1-c_{\rm tidal}) \right. \nonumber \\
 & & \left. + h_2(\xi) (1-c_{\rm tidal})^2 \left({j^\prime \over j_0}\right)^2 \left( r \over r^\prime \right)^4\right],
\end{eqnarray}
where $v^\prime = \sqrt{\G M^\prime(r^\prime)/r^\prime}$ and $j^\prime = v^\prime r^\prime$. For cosmological halos we would
expect\footnote{At the virial radius, $r$, of any halo of mass $M$ we have $v^2 = \G M/r$. Assuming that cosmological halos at any given redshift all have the same mean density within their virial radius, then $M\propto r^3$ and so $v^2 \propto r^2$.} $v\propto r$ and so $j \propto r^2$. As such, since $r\ll r^\prime$, the second term in the above will be negligible and
\begin{equation}
\Delta E = {v^{\prime 2} \over 2} \left( {r \over r^\prime} \right)^2 h_1(\xi) (1-c_{\rm tidal}).
\end{equation}

For non-circular orbits, the problem becomes more complicated. The maximum energy changes depends on the value of $\chi$ at which
apocentre and pericentre occur, and on the shape of the orbit. Experiments using elliptical orbits of eccentricity $e$ show that
using the semi-major axis in place of $r$ and
\begin{equation}
h_1(\xi,e) = h_1(\xi) (1+e)^{2.17} (1-e)^{-2.44}
\end{equation}
is a reasonable approximation (better than 20\%) for $e<0.8$. Application to the more complex orbits found in typical dark matter halos is deferred to a future paper. As such, we will ignore any dependence of $h_1(\xi)$ on orbital shape in this work.

\subsection{Coriolis Torque}

In the rotating frame, the particle experiences a Coriolis force in addition to the tidal force, which can be of comparable
magnitude. This results in a torque
\begin{eqnarray}
{\bf T}_{\rm Coriolis} & = & -2 {\bf r} \wedge ( \mathbf{\Omega} \wedge {\bf v}) \nonumber \\
 & = & -2 [ ({\bf r}\cdot{\bf v}) \mathbf{\Omega} - ({\bf r}\cdot \mathbf{\Omega}) {\bf v}].
\end{eqnarray}
For circular orbits ${\bf r}\cdot{\bf v}=0$, so
\begin{equation}
{\bf T}_{\rm Coriolis} = 2  ({\bf r}\cdot \mathbf{\Omega}) {\bf v}.
\end{equation}
Note that vector $\mathbf{\Omega}$ is normal to ${\bf
r}^\prime$. Defining angle $\gamma$ as a measure of the rotation of the normal to the orbital plane around ${\bf r}^\prime$ with
$\gamma=0$ corresponding to the plane containing ${\bf r}^\prime$ and $\mathbf{\Omega}$ we have
\begin{equation}
{\bf T}_{\rm Coriolis} = 2 r\Omega (\cos\xi\cos\gamma\cos\chi - \sin\gamma\sin\chi) {\bf v}.
\end{equation}
The change in angular momentum due to the Coriolis force is then
\begin{equation}
\Delta {\bf j}_{\rm Coriolis} =  \int 2 {r^3 \Omega \over j_0} (\cos\xi\cos\gamma\cos\chi - \sin\gamma\sin\chi) {\bf v} \d \chi.
\end{equation}
Projecting this onto our basis vectors we find
\begin{eqnarray}
\Delta {\bf j} & = & 2 \left({r \over r^\prime}\right)^2 v^\prime r^\prime \int (\cos\xi\cos\gamma\cos\chi - \sin\gamma\sin\chi) \nonumber \\
 &  & \times \left( \begin{array}{c}
\sin\chi \sin\xi \\
\cos\chi \\
\sin\chi\cos\xi
\end{array} \right)
\d \chi.
\end{eqnarray}
where we have used $v=j_0/r$ and $\Omega = v^\prime/r^\prime$. Alternatively,
\begin{eqnarray}
\Delta {\bf j} & = & 2 \left({r \over r^\prime}\right)^2 \sqrt{{g_{\rm tidal} \over 1-c_{\rm tidal}}} j_0 \int (\cos\xi\cos\gamma\cos\chi - \sin\gamma\sin\chi) \nonumber \\
 & & \times \left( \begin{array}{c}
\sin\chi \sin\xi \\
\cos\chi \\
\sin\chi\cos\xi 
\end{array} \right)
\d \chi.
\end{eqnarray}

Note that, for cosmological halos, we expect characteristic properties of halos of different masses to obey $v\propto r$. As such,
we find $g_{\rm tidal} \propto (r^\prime/r)^4$. It can then be seen that the change in angular momentum due to the tidal force and
the Coriolis force are of comparable magnitude.

We can compute the functions $h_1(\xi)$ and $h_2(\xi)$ when the Coriolis torque is included, by averaging over angle $\gamma$. The
results for $h_1(\xi)$ is shown as a dashed line in Fig.~\ref{fig:hxi}. We find that the inclusion of the Coriolis torque significantly alters $h_1(\xi)$ (by up to a factor of around 2). We therefore choose to include the Coriolis term in all calculations carried out in this work.

The case of non-circular orbits including Coriolis torques becomes significantly more complicated and a full treatment is deferred
to a future paper.

\subsection{Final Result}

The final statement is that the change in energy of the particle can be approximated by
\begin{equation}
\Delta E = {v^{\prime 2} \over 2} \left( {r \over r^\prime} \right)^2 h_1(\xi) (1-c_{\rm tidal}),
\label{eq:deltaE}
\end{equation}
where $h_1(\xi)$ is computed with the inclusion of the Coriolis torque term as shown in Fig.~\ref{fig:hxi}. Thus, the maximum energy
of a particle relative to the satellite centre is
\begin{equation}
E_{\rm max} = E_0 + \Delta E.
\end{equation}
It is this value of $E_{\rm max}$ which is used in eqn.~(\ref{eq:mloss}).


\begin{thebibliography}{}
\bibitem[Benson et al. <2002>]{benson02}Benson~A.~J., Frenk~C.~S., Lacey~C.~G., Baugh~C.~M., Cole~S., 2002, MNRAS, 333, 177	
\bibitem[Benson et al. <2005>]{benson05}Benson~A.~J., 2005, MNRAS, 358, 551
\bibitem[Bullock et al. <2000>]{bullock00}Bullock~J.~S., Kravtsov~A.~V., Weinberg~D.~H., 2000, ApJ 539, 517
\bibitem[Bullock et al. <2001>]{bullock01}Bullock~J.~S., Kolatt~T.~S., Sigad~Y., Somerville~R.~S., Kravtsov~A.~V., Klypin~A.~A., Primack~J.~R., Dekel~A., 2001, MNRAS, 321, 559
\bibitem[Calc\'aneo-Rold\'an et al. <2000>]{calcaneo00}Calc\'aneo-Roldan~C., Moore~B., Bland-Hawthorn~J., Malin~D., Sadler~E.~M., 2000, MNRAS, 314, 324
\bibitem[Just \&  Penarrubia <2004>]{just04}Just~A., Penarrubia~J., 2004, A\&A 431, 861
\bibitem[Hayashi et al. <2003>]{hayashi03} Hayashi~E., Navarro~J.~F., Taylor~J.~E., Stadel~J., Quinn~T., 2003, ApJ, 584, 541
\bibitem[Mateo <1998>]{mateo98}Mateo~M., 1998, ARA{\&}A, 36, 435
\bibitem[Navarro, Frenk \& White <1995>]{navarro95}Navarro~J.~F., Frenk~C.~S., White~S.~D.~M., 1995, MNRAS, 275,  56
\bibitem[Moore et al. <1999>]{moore99} Moore~B., Quinn~T., Governato~F., Stadel~J., Lake~G., 1999, MNRAS, 310, 1147
\bibitem[Ibata et al. <1994>]{ibata94}Ibata~R.~A., Gilmore~G., Irwin, 1994, Nature, 370, 194
\bibitem[Power et al. <2003>]{poweretal}Power~C., Navarro~J.~F., Jenkins~A., Frenk~C.~S., White~S.~D.~M., Springel~V., Stadel~J., Quinn~T., 2003, MNRAS, 338, 14
\bibitem[Odenkirhen et al. <2001>]{odenkirhen01}Odenkirhen~M. et al., 2001, ApJ, 548, L165
\bibitem[Penarrubia et al. <2004>]{penarrubia04}Penarrubia~J., Just~A., Kroupa~P., 2004, MNRAS, 349, 747
\bibitem[Read et al. <2006>]{read06}Read~J.~I., Wilkinson~M.~I., Evans~N.~W., Gilmore~G., Kleyna~J.~T. 2006, MNRAS, 366, 429
\bibitem[Springel <2005>]{gadget2}Springel~V., 2005, MNRAS, 364, 1105
\bibitem[Tasitsiomi <2003>]{tasitsiomi02}Tasitsiomi~A., 2003, Int.~J.~Mod.~Phys., D12, 1157
\bibitem[Taylor \& Babul <2001>]{taylor01}Taylor~J.~E., Babul~A., 2001, ApJ, 559, 716
\bibitem[Zwicky <1951>]{zwicky51}Zwicky~F., 1951, PASP, 63, 61
 
\end{thebibliography}
\end{document}